# Temporal Effective Medium for Programmable Acoustic Metamaterials with Multiple Resonances

Xinghong Zhu[1], Hong-Wei Wu[3*], and Jensen Li[1,2*]

[1]Department of Physics, The Hong Kong University of Science and Technology, Clear Water Bay, Hong Kong, China
[2]Department of Engineering, University of Exeter, EX4 4QF, United Kingdom
[3]School of Mechanics and Photoelectric Physics, Anhui University of Science and Technology, Huainan 232001, China

We extend effective medium theory (EMT) to time-modulated, frequency-dispersive acoustic metamaterials with multiple resonances. While previous studies focused on non-dispersive or single-resonance systems, advances in programmable materials now enable precise control of time-varying responses. We derive explicit averaging rules that account for the interplay between resonant and modulation frequencies. When resonant frequencies are much lower than the modulation frequency, modulating the resonant strength yields the temporal average of monopolar susceptibility $\chi$, while modulating the resonant frequency results in the average of $1/\chi$, applied per resonance mode. In hybrid cases, high-frequency resonances (relative to modulation) can be renormalized as a non-dispersive background before averaging the rest. This generalized temporal EMT offers a unified framework for designing compact, topologically robust, and non-Hermitian acoustic devices, leveraging the possible programmability of time-dependent material parameters in future.

* Emails: j.li13@exeter.ac.uk; hwwu@aust.edu.cn

**INTRODUCTION**

Metamaterials, built from periodic meta-atoms, have enabled unprecedented wave manipulation, including negative refraction, superlensing, and perfect absorption [1-4]. A key concept is the use of effective medium theory (EMT) [5-12], which describes subwavelength structures using macroscopic parameters. EMT supports a two-step design process: first, defining the desired effective medium profile; then, engineering the metamaterial to realize it by either numerical effective medium extraction or formulas. This approach has enabled many exotic wave phenomena including invisibility cloak, super resolution imaging, wave tunneling, amplification, sensing and beyond [13-19]. Recently metamaterials have been extended to the time-varying regime, where time-varying material parameters are used as a new degree of freedom for wave control. Time-varying metamaterials enable four-dimensional modulation (x, y, z, t) [20-22], giving rise to effects like temporal refraction, frequency conversion, nonreciprocity, and momentum bandgaps [23-39]. These developments have spurred extensions of EMT into the temporal domain, typically requiring modulation frequencies much higher than the signal frequency. Existing formulations, however, mostly address non-dispersive or single-resonance systems [40-43].

A promising route for a fast temporal modulation in acoustics uses meta-atoms with digital feedback circuits, enabling real-time tunability [29,44]. This method also supports modulation of generalized resonance types. Similar to how complex electromagnetic dispersion can be modeled by multiple Lorentzian resonances, digital feedback allows analogous implementations in time-varying systems. Although single-resonance temporal EMT has been experimentally demonstrated [45,46], a general framework for multi-resonance systems remains undeveloped.

In this work, we extend the temporal effective medium theory to encompass frequency-dispersive metamaterials with multiple resonances, focusing on acoustic systems for clarity and demonstration. By modulating different resonance parameters, such as resonant frequency and strength, we derive effective medium formulas that account for the interplay between resonant frequencies and modulation frequency. Our results reveal that when resonant frequencies are much smaller than the modulation frequency, modulating resonant strength follows temporal averaging of monopolar susceptibility directly, while modulating resonant frequency inverts this relationship, producing reciprocal averaging, applied separately to each resonance mode. These principles enable precise control over effective responses, even in hybrid regimes where some resonant frequencies fall in the non-dispersive regime (much higher than the modulation frequency), these modes will suppress dispersion in their resonances, forming an effectively static, renormalized background. This allows the dispersive component to be time-averaged with the normalized background. These findings provide a critical foundation for understanding and designing temporal metamaterials with fast modulation, enabling compact, topologically robust, and non-Hermitian acoustic devices with tailored functionalities achievable through time-varying material parameters in future. Furthermore, the principles developed here can be extended to other wave systems, opening new possibilities for programmable metamaterial design.

**RESULTS**

**Modulating resonant strength with two resonances**

We begin with a one-dimensional (1D) spatially homogeneous acoustic medium along $x$, where the material properties are modulated in time $t$ to generate monopolar polarization $M(x,t)$ in the surrounding air. The acoustic wave propagation in this medium is governed by the following equations:

$$\partial_x p(x,t) + \rho_0 \partial_t v(x,t) = 0, \tag{1}$$

$$\partial_x v(x,t) + \beta_0 \partial_t (p(x,t) + M(x,t)) = 0, \tag{2}$$

where $\rho_0$ and $\beta_0$ are the density and compressibility of air, and $p(x,t)$ and $v(x,t)$ are the pressure and velocity fields, respectively. We assume the metamaterial possesses only monopolar responses, composed of several resonant modes by $M(x,t) = \Sigma_i M_i(x,t)$ and generate monopolar susceptibility $\chi = \Sigma_i \chi_i$. Each monopolar response $M_i(x,t)$ corresponds to the $i$-th resonant mode and interacts with the pressure field $p(x,t)$ according to a Lorentzian-type model:

$$\partial_t^2 M_i(x,t) + 2\gamma_i(t)\partial_t M_i(x,t) + \omega_{0i}^2(t) M_i(x,t) = a_i(t) p(x,t), \tag{3}$$

where $a_i(t)$, $\omega_{0i}(t)$ and $\gamma_i(t)$ the time-dependent resonant strength (with units of squared frequency), resonant frequency, and linewidth of the $i$-th resonance, respectively. As shown in Fig. 1(a), the three resonance parameters $(a_i, \omega_{0i}, \gamma_i)$ are modulated periodically with period $T$, consisting of two phases. Phase A lasts for $\xi T$ (duty cycle $\xi$), during which the parameters remain constant as $(a_{iA}, \omega_{0iA}, \gamma_{iA})$. Phase B spans the remaining $(1 - \xi)T$ with parameters $(a_{iB}, \omega_{0iB}, \gamma_{iB})$. For clarity, we focus on monopolar polarizations with two resonances ($i = 1,2$), while cases with more resonances can be derived from the findings for two resonances. Figure 1(b) schematically depicts the static compressibility $\beta_A$ and $\beta_B$ ($\beta = 1 + \chi_1 + \chi_2$), each featuring two resonances in phases A and B, assuming no time modulation. Our goal is to derive the temporal effective medium formula for modulating these phases with duty cycle $\xi$ in terms of $\beta_A$ and $\beta_B$. The formulas vary significantly depending on which resonance parameters (e.g., strength or resonant frequency) or their combinations are modulated, and whether the non-dispersive limit is approached by having the resonant frequency $\omega_0$ much larger than the modulation frequency $1/T$.

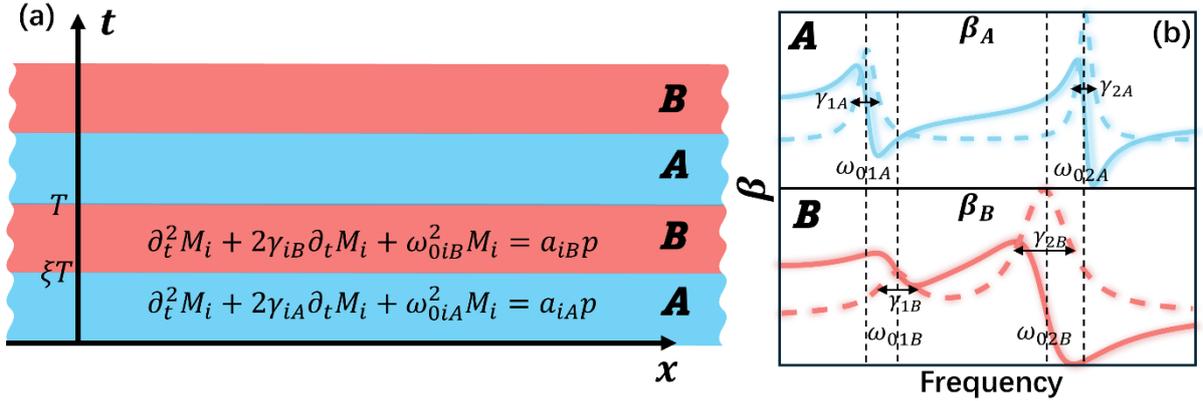

Fig. 1 Schematic diagram of a time-varying metamaterial with multiple resonance modes. (a) Each resonance $i$ has a monopolar polarization $M_i$ driven by pressure $p$, governed by Lorentzian-type differential equations with resonant frequency, strength and linewidth switching phases A and B, with modulation period $T$ and duty cycle $\xi: 1 - \xi$. (b) Static effective compressibility $\beta_A$ and $\beta_B$ in phases A (blue) and B (red), assuming no time modulation, with $\omega_{0iA}, \omega_{0iB}$ and $\gamma_{iA}, \gamma_{iB}$ as the resonant frequencies and linewidths for resonance $i$, respectively. Solid (dashed) lines represent the real (imaginary) parts.

Firstly, we consider the case where only the resonant strengths $a_1(t)$ and $a_2(t)$ are time modulated, while the other parameters $\omega_{0i}$ and $\gamma_i$ remain constant, as shown in Fig. 2(a). To derive the temporal effective medium, we assume the medium is homogenous and infinite along the $x$ direction, ensuring that the propagation constant $k$ remains unchanged across different time interfaces. Substituting $\partial_x$ with $ik$, Eq. (1)-(3) can be rewritten as

$$i\partial_t \psi = \widehat{\omega}\psi, \quad \widehat{\omega} = \begin{pmatrix} 0 & k/\beta_0 & 0 & i & 0 & i \\ k/\rho_0 & 0 & 0 & 0 & 0 & 0 \\ 0 & 0 & 0 & -i & 0 & 0 \\ -ia_1(t) & 0 & i\omega_{01}^2 & -2i\gamma_1 & 0 & 0 \\ 0 & 0 & 0 & 0 & 0 & -i \\ -ia_2(t) & 0 & 0 & 0 & i\omega_{02}^2 & -2i\gamma_2 \end{pmatrix}, \quad (4)$$

with the eigenstate vector $\psi = (p, v, M_1, -\partial_t M_1, M_2, -\partial_t M_2)^T$. The propagation matrix $\widehat{\omega}$, alternates between $\widehat{\omega}_A$ and $\widehat{\omega}_B$, corresponding to resonant strength $a_{1A}$, $a_{2A}$ and $a_{1B}$, $a_{2B}$ respectively. In the two static cases, Eq. (1)-(3) can be solved time-harmonically and combined with the constitutive relationship $p + M = \beta p = (1 + \chi_1 + \chi_2)p$ with $M = M_1 + M_2$, where $\beta$ is the effective compressibility and $\chi_i$ is the effective susceptibility relative to air for the $i$-th resonance. This yields

$$\beta_{A/B} = 1 + \chi_{1A/B} + \chi_{2A/B} = 1 + \frac{a_{1A/B}}{\omega_{01}^2 - \omega^2 - 2i\omega\gamma_1} + \frac{a_{2A/B}}{\omega_{02}^2 - \omega^2 - 2i\omega\gamma_2}. \quad (5)$$

An example of the compressibility for two static phases is shown in Fig. 2(b), with blue and red lines representing phases A and B, respectively, and solid/dashed lines indicating the real/imaginary parts. For phase A (blue), the resonant frequencies are set as $\omega_{01}/(2\pi) = $ 1kHz, $\omega_{02}/(2\pi) = 1.5$kHz, with linewidths $\gamma_1/\omega_{01} = \gamma_2/\omega_{02} = 0.05$. The resonant strengths are $a_{1A}/\omega_{01}^2 = a_{2A}/\omega_{02}^2 = 0.2$. In phase B (red), the resonant strengths change to $a_{1B}/\omega_{01}^2 = 0.5$ and $a_{2B}/\omega_{02}^2 = 0.6$, while all other parameters remain constant. In the time-varying case, the resonant strengths switch between $a_{iA}$ and $a_{iB}$. Assuming the modulation period $T$ satisfies $\omega T \ll 1$ and $\omega_{0i} T \ll 1$, the modulation frequency is much higher than both the signal frequency (or eigenfrequency in the medium) and the resonant frequencies. Under these conditions, the total transfer matrix $\exp(-i\widehat{\omega}_B(1-\xi)T)\exp(-i\widehat{\omega}_A T)$ can be approximated as $\exp(-i\widehat{\omega}_{\text{eff}} T)$, where $\widehat{\omega}_{\text{eff}} = \xi\widehat{\omega}_A + (1-\xi)\widehat{\omega}_B$ in the first order Taylor expansion of $T$. Using Eq. (5), the effective compressibility for modulating $a_i$ is given by

$$\beta_{\text{eff}} = 1 + \chi_{1\text{eff}} + \chi_{2\text{eff}}. \quad (6)$$

$$\chi_{1\text{eff}} = \xi\chi_{1A} + (1-\xi)\chi_{1B}, \quad \chi_{2\text{eff}} = \xi\chi_{2A} + (1-\xi)\chi_{2B}. \quad (7)$$

The result corresponds to the temporal average of compressibility $\beta$ or, equivalently, the susceptibility $\chi$.

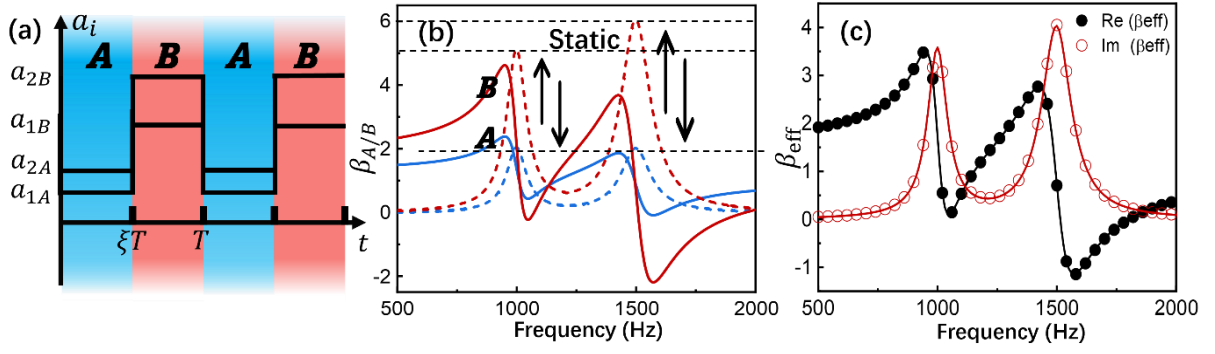

Fig. 2 (a) Sketch of time-varying metamaterials with two monopolar resonant strengths modulated over period $T$ and duty cycle $\xi$. (b) Static compressibility $\beta_A$ (blue) in phase A and $\beta_B$ (red) in phase B, with solid (dashed) line representing real (imaginary) parts. The parameters are chosen as $\omega_{01}/(2\pi) = 1$ kHz, $\omega_{02}/(2\pi) = 1.5$ kHz, $a_{1A}/\omega_{01}^2 = a_{2A}/\omega_{02}^2 = 0.2$, $a_{1B}/\omega_{01}^2 = 0.5$, $a_{2B}/\omega_{02}^2 = 0.6$, $\gamma_1/\omega_{01} = \gamma_2/\omega_{02} = 0.05$. (c) Effective compressibility $\beta_{\text{eff}}$ extracted using the eigenmode approach (red and black dots), with $a_1(t)$ switching between $a_{1A}$ and $a_{1B}$ and $a_2(t)$ switching between $a_{2A}$ and $a_{2B}$ ($\xi = 0.5$). Analytic results (red and black lines) are from Eq. (6) and (7). Modulation frequency: $1/T = 8$kHz; other parameters as in (b).

Alternatively, the eigenmode of Eq.(4) can be numerically solved using the transfer matrix method. At a small eigenfrequency $\omega$, we can extract the effective compressibility $\beta_{\text{eff}}$ and density $\rho_{\text{eff}}$ (relative to air) from the eigenstate as [46]

$$\beta_{\text{eff}} = \frac{k}{\omega \rho_0} \frac{p + M}{v},$$

$$\rho_{\text{eff}} = \frac{k}{\omega \beta_0} \frac{v}{p + M}. \quad (8)$$

This approach is employed to generate all numerical results in the following. As an example, we modulate resonant strength $a_i(t)$ with the modulation parameters used in Fig. 2(b) with duty cycle $\xi = 0.5$ and calculated the effective compressibility $\beta_{\text{eff}}$ marked as black and red hollow dots for real and imaginary part against operational frequency $\omega/(2\pi)$ in Fig. 2(c). These results match well with the effective formula of time averaging $\chi_i$ for each mode by Eq.(7) and thus obtaining the effective compressibility by Eq. (6), as plot in black and red lines.

**Modulating resonant frequency with two resonances**

Following the case of modulating resonant strength $a_i$, we now consider modulating resonant frequency $\omega_{0i}$ while keeping $a_i$ and $\gamma_i$ constant, as shown in Fig. 3(a). $\omega_{0i}$ switches between $\omega_{0iA}$ and $\omega_{0iB}$ in phases A and B, respectively. We still require $\omega_{0iA/B}T \ll 1$ in both phases. This allows us to approximate the effective propagation matrix using the lowest-order Talor expansion: $\widehat{\omega}_{\text{eff}} = \xi\widehat{\omega}_A + (1-\xi)\widehat{\omega}_B$. The temporal effective medium formula in modulating $\omega_{0i}$ can then be expressed in terms of the static susceptibilities as

$$1/\chi_{1\text{eff}} = \xi/\chi_{1A} + (1-\xi)/\chi_{1B}, \quad 1/\chi_{2\text{eff}} = \xi/\chi_{2A} + (1-\xi)1/\chi_{2B}, \tag{9}$$

where the static monopolar susceptibility in each phase is

$$\chi_{1A/B} = \frac{a_1}{\omega_{01A/B}^2 - \omega^2 - 2i\omega\gamma_1}, \quad \chi_{2A/B} = \frac{a_2}{\omega_{02A/B}^2 - \omega^2 - 2i\omega\gamma_2}. \tag{10}$$

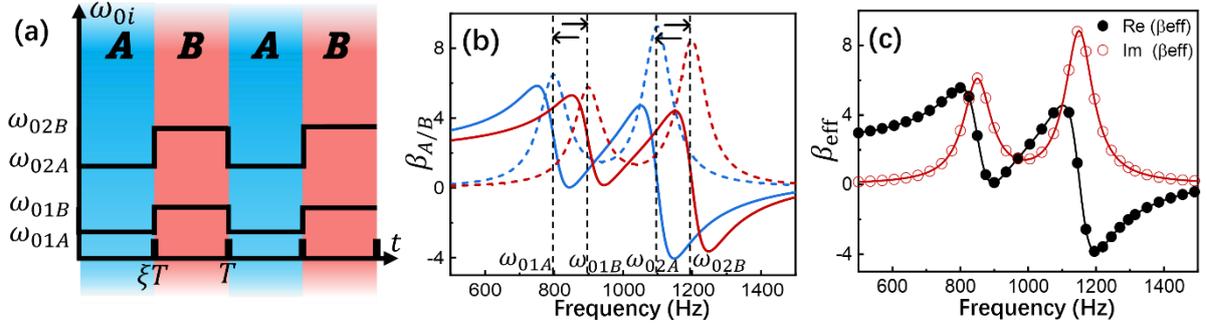

Fig. 3 Time-varying metamaterial with modulated resonant frequencies. (a) Schematic diagram of metamaterial with resonant frequency $\omega_{0i}$ switching between $\omega_{0iA}$ and $\omega_{0iB}$. (b) Static compressibility in phase A (blue) with $\omega_{01A}/(2\pi) = 0.8$ kHz, $\omega_{02A}/(2\pi) = 1.1$ kHz and in phase B (red) with $\omega_{01B}/(2\pi) = 0.9$ kHz, $\omega_{02B}/(2\pi) = 1.2$ kHz. (c) Effective compressibility when modulating resonant frequency according to (b) with duty cycle $\xi = 0.5$, with dots from the eigenmode approach (Eq. (8)) and lines from Eq. (9). The resonant strength and linewidth are both constants: $a_1/(2\pi)^2 = 0.5$ kHz$^2$, $a_2/(2\pi)^2 = 1$ kHz$^2$, $\gamma_1/(2\pi) = \gamma_2/(2\pi) = 50$ Hz.

Figure 3(b) shows the static compressibility for phase A (blue) and phase B (red) as a function of operational frequency $\omega/(2\pi)$, with solid and dashed lines representing real and imaginary parts, respectively. The resonant frequencies are $\omega_{01A}/(2\pi) = 0.8$ kHz and $\omega_{02A}/(2\pi) = 1.1$ kHz for phase A, and $\omega_{01B}/(2\pi) = 0.9$ kHz and $\omega_{02B}/(2\pi) = 1.2$ kHz for phase B. The resonant strengths and linewidths remain constant: $a_1/(2\pi)^2 = 0.5$ kHz$^2$, $a_2/(2\pi)^2 = 1$ kHz$^2$, $\gamma_1/(2\pi) = \gamma_2/(2\pi) = 50$ Hz. After defining the static compressibilities for the two phases, they are alternated with a duty cycle of $\xi = 0.5$. The effective compressibility $\beta_{\text{eff}}$ is calculated using the eigenmode approach and shown in Figure 3(c) as black and red hollow dots for the real and imaginary parts. These results align with the analytic predictions (solid lines) obtained from Eq. (9), confirming the accuracy of the time-averaging approach for $1/\chi_i$.

**Effective medium formulas with $\omega_{01}T \ll 1$ and $\omega_{02}T \gg 1$**

Thus far, we have presented modulation cases satisfying both $\omega T \ll 1$ and $\omega_{0i}T \ll 1$. While $\omega T \ll 1$ is a necessary condition for the temporal effective medium criteria, $\omega_{0i}T \ll 1$ is not. When both resonant frequencies $\omega_{01}$ and $\omega_{02}$ are sufficiently large ($\omega_{0i}T \gg 1$), the system approaches the so-called frequency-nondispersive regime. In this case, modulating $a_i$ allows the susceptibilities to be expressed as $\chi_{1A/B} = a_{1A/B}/\omega_{01}^2$ and $\chi_{2A/B} = a_{2A/B}/\omega_{02}^2$ with $a_i$ scaling with $\omega_0^2$ to have non-zero $\chi_i$. Although $\omega T \ll 1$ still holds, the large $\omega_{0i}T$ prevents the direct application of the first-order Taylor expansion. However, if the resonant linewidth $\gamma_i$ is sufficiently large, it can supress the rapid oscillation of monopolar polarization $M_i$ triggered at each time boundary and approach the new static regime. By examining Eq. (4) for large $\omega_{0i}$, the wave equation in phase A/B can be approximated as

$$i\partial_t(p+M) \cong \frac{k}{\beta_0}v, \quad i\partial_t v \cong \frac{k}{\rho_0}\frac{1}{1+\chi_{1A/B}+\chi_{2A/B}}(p+M), \quad (11)$$

which effectively reduces the propagation matrix $\widehat{\omega}$ in Eq. (4) from 6×6 to 2×2 and yields the temporal effective formula for time-varying modulation amplitude $a_i(t)$ at large $\omega_{0i}$:

$$\frac{1}{1+\chi_{1\text{eff}}+\chi_{2\text{eff}}} = \frac{\xi}{1+\chi_{1A}+\chi_{2A}} + \frac{1-\xi}{1+\chi_{1B}+\chi_{2B}}. \quad (12)$$

This result ($\beta_{\text{eff}} = 1 + \chi_{1\text{eff}} + \chi_{2\text{eff}}$) corresponds to the temporal average of $1/\beta$, as first investigated by Engheta [41] for non-dispersive time-varying media.

After recognizing that effective medium formulas can be influenced by the value of $\omega_{0i}$, we consider a hybrid case where $\omega_{01}T \ll 1$ and $\omega_{02}T \gg 1$. Under this condition, only $\chi_{2A/B}$ can be expressed as $\chi_{2A/B} = a_{2A/B}/\omega_{02}^2$ through modulation of $a_i$. Together with Eqs. (1)-(3), the original 6×6 eigenvalue problem (Eq. (4)) can be truncated to a 4 by 4 system:

$$i\partial_t\psi = \widehat{\omega}\psi, \quad \widehat{\omega} = \begin{pmatrix} 0 & k/\beta_0 & 0 & i \\ \dfrac{k/\rho_0}{1+\chi_2(t)} & 0 & 0 & 0 \\ 0 & 0 & 0 & -i \\ -\dfrac{ia_1(t)}{1+\chi_2(t)} & 0 & i\omega_{01}^2 & -2i\gamma_1 \end{pmatrix}, \quad (13)$$

where now the state vector becomes $\psi = (p + M_2, v, M_1, -\partial_t M_1)^T$. The propagation matrix $\widehat{\omega}_A$ and $\widehat{\omega}_B$ correspond to the parameter pairs $(a_{1A}, \chi_{2A})$ and $(a_{1B}, \chi_{2B})$, respectively. The total transfer matrix $\exp(-i\widehat{\omega}_B(1-\xi)T)\exp(-i\widehat{\omega}_A T)$ can be approximated by $\exp(-i\widehat{\omega}_{\text{eff}}T)$ where $\widehat{\omega}_{\text{eff}} = \xi\widehat{\omega}_A + (1-\xi)\widehat{\omega}_B$, using the first-order Taylor expansion. In this case, the effective $\chi_{i\text{eff}}$ can be obtained as

$$\frac{1}{1+\chi_{2\text{eff}}} = \frac{\xi}{1+\chi_{2A}} + \frac{1-\xi}{1+\chi_{2B}},$$

$$\frac{\chi_{1\text{eff}}}{1+\chi_{2\text{eff}}} = \frac{\xi\chi_{1A}}{1+\chi_{2A}} + \frac{(1-\xi)\chi_{1B}}{1+\chi_{2B}}. \quad (14)$$

As indicated by the (2,1) entry of $\widehat{\omega}$, the nondispersive component $\chi_{2\text{eff}}$ is independently determined by the temporal average of $1/(1+\chi_2)$, denoted as $\langle 1/(1+\chi_2)\rangle$, where "$\langle\rangle$" represents time averaging. In the absence of $\chi_1$, this aligns with the temporal average of $1/\beta$ for nondispersive systems. For the dispersive component, Eq. (7) remains valid for the temporal averaging of susceptibility $\langle\chi\rangle$ but now has to be normalized by the nondispersive term ($1+\chi_2$). Specifically, the time-averaged $\langle\chi_1/(1+\chi_2)\rangle$ yields $\chi_{1\text{eff}}/(1+\chi_{2\text{eff}})$, consistent with the (4,1) entry of $\widehat{\omega}$. Here, $\chi_{2\text{eff}}$ can be regarded as the background susceptibility (zero in air, assuming no $\chi_2$ contribution). The total effective compressibility then follows $\beta_{\text{eff}} = 1 + \chi_{1\text{eff}} + \chi_{2\text{eff}}$.

Fig. 4 demonstrates the phase transition resulting from modulating the resonant strength $a_i$ while keeping resonant frequency $\omega_{01}$ fixed and gradually increasing $\omega_{02}$. The schematic representation is shown in Fig. 4(a). Initially, when both $\omega_{01}$ and $\omega_{02}$ are small ($\omega_{0i}T \ll 1$), the effective compressibility $\beta_{\text{eff}}$ corresponds to the temporal average of $\beta$ (equivalent $\chi$), as shown in the left panel of Fig. 4(b). A phase transition occurs when resonant frequency $\omega_{02}$ approaches the modulation frequency $1/T = 8\text{kHz}$. As $\omega_{02}$ continues to increase, $\beta_{\text{eff}}$ asymptotically converges to the high frequency limit 1.85, as calculated from Eq. (14) (green

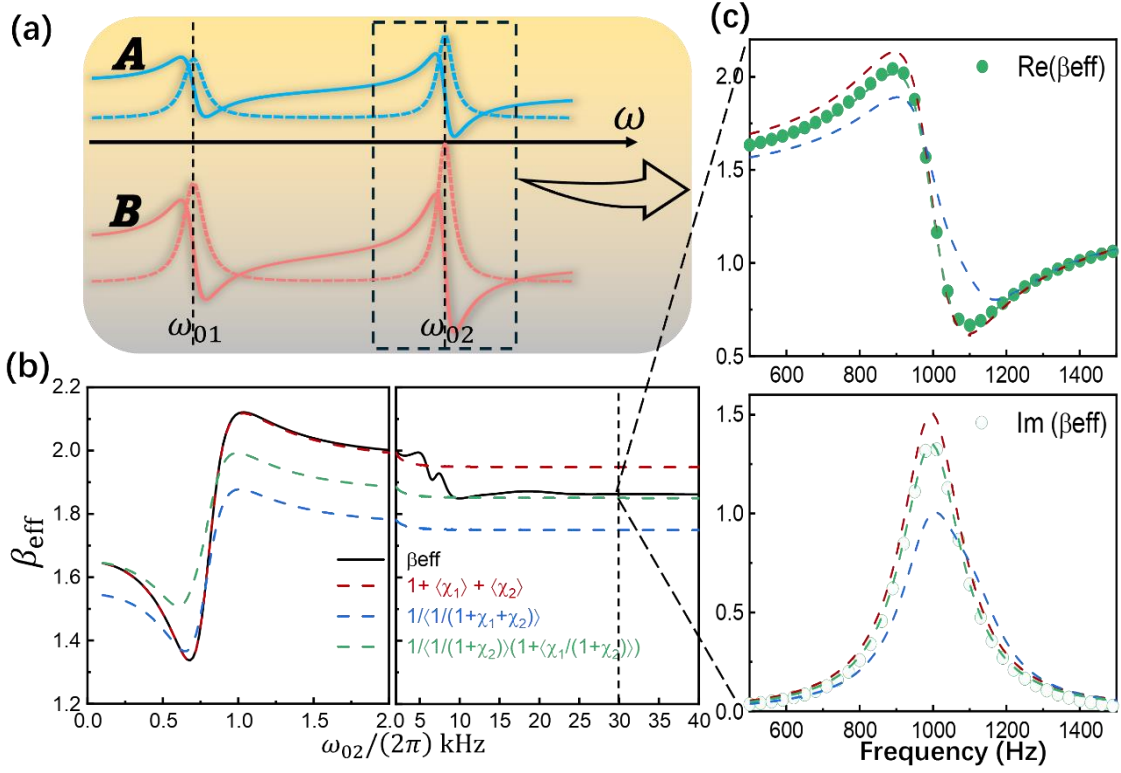

Fig. 4 (a) Conceptual picture of modulating resonant strengths with one resonance ($\omega_{02}$) shifting to high frequencies. (b) Phase transition of a temporal metamaterial with time modulated resonant strengths $a_i(t)$ as $\omega_{02}$ increases while fixing $\omega_{01}/(2\pi) = 1\text{kHz}$. The black solid line shows the effective compressibility ($\beta_{\text{eff}}$) from numerical eigenmode calculations. Red and blue dashed lines represent the temporal average of compressibility ($\beta$) and bulk modulus ($1/\beta$), respectively. The green dashed line is the averaging formula from Eq. (14). Resonant strengths alternate between $a_{1A} = 0.1\omega_{01}^2$, $a_{1B} = 0.5\omega_{01}^2$ for $a_1(t)$ and $a_{2A} = 0.1\omega_{02}^2$, $a_{2B} = 0.5\omega_{02}^2$ for $a_2(t)$ with $\xi = 0.5$. The operational frequency and linewidths are $\omega/(2\pi) = 0.8\text{kHz}$, $\gamma_1/\omega_{01} = 0.1$ and $\gamma_2/\omega_{02} = 0.2$. (c) Effective compressibility at varying operational frequencies with modulated resonant strengths $a_i(t)$ at fixed $\omega_{02}/(2\pi) = 30\text{kHz}$ with other parameters in (b). Green solid and hollow dots represent the real and imaginary part obtained from eigenmode approach. The red, green and dashed lines follow the same representations as in (b).

dashed line), shown in the right panel of Fig. 4(b). The effective compressibility $\beta_{\text{eff}}$ obtained from eigenmode approach (black solid lines) agrees with the analytic results for both cases ($\omega_{02}T \ll 1$ and $\omega_{02}T \gg 1$). The operational frequency is fixed at 0.8kHz and detailed parameters used are listed in the caption. To further validate Eq. (14), Fig. 4(c) shows a scan of the operational frequency from 0.5kHz to 1.5kHz, with $\omega_{02}/(2\pi)$ fixed at 30kHz. The eigenmode-extracted effective compressibility (green dots) exhibits excellent agreement with the analytic predictions (green dashed lines) for both real and imaginary parts, demonstrating the robustness of our temporal effective medium formula across operational frequencies.

Finally, we shift to modulating resonant frequencies $\omega_{0i}$ with small $\omega_{01}$ ($\omega_{01A/B}T \ll 1$) but large $\omega_{02}$ ($\omega_{02A/B}T \gg 1$), as schematically shown in Fig. 5(a). In this case, we adopt the same approach as Eq. (13) to approximate $\widehat{\omega}_{\text{eff}}$ but now $\widehat{\omega}_A$ and $\widehat{\omega}_B$ correspond to resonant frequency pairs $(\omega_{01A}, \omega_{02A})$ and $(\omega_{01B}, \omega_{02B})$, respectively. The effective medium formula can be deduced as:

$$\frac{1}{1+\chi_{2\text{eff}}} = \frac{\xi}{1+\chi_{2A}} + \frac{1-\xi}{1+\chi_{2B}},$$

$$\frac{1}{\chi_{1\text{eff}}} = \frac{\xi}{\chi_{1A}} + \frac{(1-\xi)}{\chi_{1B}}.$$

(15)

Similar to the case of modulating resonant strength, the nondispersive part $\chi_2$, is still governed by the time average of $1/(1+\chi_2)$. For the dispersive component $\chi_1$, however, the modulation now affects the resonant frequency, as reflected in the (4,3) entry of $\widehat{\omega}$ in Eq.(13). This requires applying Eq. (9) to the temporal average of $1/\chi_1$ to obtain $\chi_{1\text{eff}}$. As an example, we modulate $\omega_{01}$ between $\omega_{01A}/(2\pi) = 0.8$kHz and $\omega_{01B}/(2\pi) = 1.2$kHz and modulate $\omega_{02}$ between $\omega_{02A}/(2\pi) = 20$kHz and $\omega_{02B}/(2\pi) = 25$kHz with duty cycle $\xi = 0.5$. Fig. 5 (a) Schematic of modulating resonant frequencies with $\omega_{01A/B}T \ll 1$ and $\omega_{02A/B}T \gg 1$. Real (b) and imaginary (c) part of effective compressibility of time-varying metamaterial with $\omega_{01}$ switching between $\omega_{01A}/(2\pi) = 0.8$kHz and $\omega_{01B}/(2\pi) = 1.2$kHz and $\omega_{02}$ alternating between $\omega_{02A}/(2\pi) = 20$kHz and $\omega_{02B}/(2\pi) = 25$kHz with duty cycle $\xi = 0.5$. The other parameters are constant: $a_1/(2\pi)^2 = 0.5$ kHz$^2$, $a_2/(2\pi)^2 = 1$kHz$^2$, $\gamma_1/(2\pi) = \gamma_2/(2\pi) = 50$Hz. (b) and (c) show the real and imaginary part of the extracted effective compressibility $\beta_{\text{eff}}$ by eigenmode approach, represented by orange solid and hollow dots. These results agree well with the analytic predictions from Eq. (15), shown as orange dashed line. For comparison, we also plot the time average of $\beta$ and $1/\beta$ as red and blue lines, both showing up two resonant peaks.

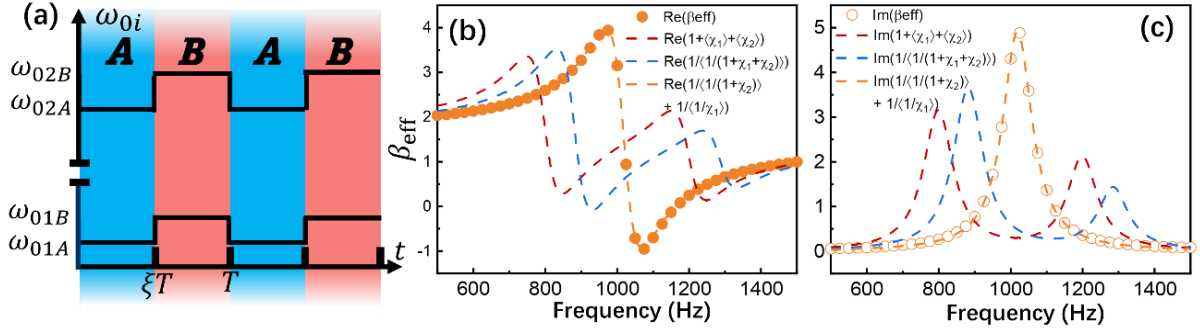

Fig. 5 (a) Schematic of modulating resonant frequencies with $\omega_{01A/B}T \ll 1$ and $\omega_{02A/B}T \gg 1$. Real (b) and imaginary (c) part of effective compressibility of time-varying metamaterial with $\omega_{01}$ switching between $\omega_{01A}/(2\pi) = 0.8$kHz and $\omega_{01B}/(2\pi) = 1.2$kHz and $\omega_{02}$ alternating between $\omega_{02A}/(2\pi) = 20$kHz and $\omega_{02B}/(2\pi) = 25$kHz with duty cycle $\xi = 0.5$. The other parameters are constant: $a_1/(2\pi)^2 = 0.5$ kHz$^2$, $a_2/(2\pi)^2 = 1$kHz$^2$, $\gamma_1/(2\pi) = \gamma_2/(2\pi) = 50$Hz.

**DISCUSSION**

In summary, we have established a comprehensive framework for temporal effective medium theory in time-modulated, frequency-dispersive acoustic metamaterials with multiple resonances. Our findings, summarized in Table 1, reveal three key rules governed by the interplay between resonant frequencies ($\omega_{0i}$) and modulation frequency ($1/T$). In the low resonant frequency regime, where both resonant frequencies are much smaller than the modulation frequency ($\omega_{0i}T \ll 1$), the effective compressibility $\beta_{\text{eff}}$ depends on whether the resonant strength $a_i$ or resonant frequency $\omega_{0i}$ is modulated: temporal average of monopolar susceptibility $\chi_i$, denoted as $\langle\chi_i\rangle$ for $a_i$ modulation and $\langle 1/\chi_i\rangle$ for modulating $\omega_{0i}$, with $\beta_{\text{eff}}$ emerging as the sum of all contributions of $\chi_{i\text{eff}}$, following $\beta_{\text{eff}} = 1 + \chi_{1\text{eff}} + \chi_{2\text{eff}}$. On the other hand, when both resonant frequencies are large ($\omega_{0i}T \gg 1$), the system falls into the nondispersive regime, giving rise to the temporal average of $1/(1 + \chi_1 + \chi_2)$. Specially, when we have a hybrid case where $\omega_{01}T \ll 1$ and $\omega_{02}T \gg 1$, large $\omega_{02}$ will render $\chi_2$ nondispersive, making $1 + \chi_2$ act as a renormalized background. In this case, we first normalize this background via the temporal average of $1/(1 + \chi_2)$, after which the dispersive part $\chi_1$ is treated by either averaging $\chi_1$ but with a scaling factor $1/(1 + \chi_2)$, i.e. $\langle\chi_1/(1 + \chi_2)\rangle$ for modulating resonant strength $a_1$ or averaging $1/\chi_1$ for modulating resonant frequency $\omega_{01}$, in accordance with the rule for the low resonant frequency case. Although our analysis focuses on systems with two resonances, the same principles can be extended to acoustic metamaterials with three or more resonances by using the same systematic approach. Beyond acoustics, these findings can be generalized to electromagnetic, elastic, and other classical wave systems, bridging temporal modulation strategies with programmable metamaterial design.

| MODULATION PARAMETERS | $\omega_{0i}T \ll 1(\omega_{0i}\downarrow)$ OR $\omega_{0i}T \gg 1(\omega_{0i}\uparrow)$ | AVERAGE METHOD |
|---|---|---|
| $a_1, a_2$ | $\omega_{01}\downarrow, \omega_{02}\downarrow$ | $\chi_{1\text{eff}} = \langle \chi_1 \rangle$ <br> $\chi_{2\text{eff}} = \langle \chi_2 \rangle$ |
| $a_1, \omega_{02}$ | $\omega_{01}\downarrow, \omega_{02}\downarrow$ | $\chi_{1\text{eff}} = \langle \chi_1 \rangle$ <br> $1/\chi_{2\text{eff}} = \langle 1/\chi_2 \rangle$ |
| $\omega_{01}, \omega_{02}$ | $\omega_{01}\downarrow, \omega_{02}\downarrow$ | $1/\chi_{1\text{eff}} = \langle 1/\chi_1 \rangle$ <br> $1/\chi_{2\text{eff}} = \langle 1/\chi_2 \rangle$ |
| $a_1, a_2$ | $\omega_{01}\downarrow, \omega_{02}\uparrow$ | $\dfrac{\chi_{1\text{eff}}}{1+\chi_{2\text{eff}}} = \langle \dfrac{\chi_1}{1+\chi_2} \rangle$ <br> $\dfrac{1}{1+\chi_{2\text{eff}}} = \langle \dfrac{1}{1+\chi_2} \rangle$ |
| $a_1, \omega_{02}$ | $\omega_{01}\downarrow, \omega_{02}\uparrow$ | $\dfrac{\chi_{1\text{eff}}}{1+\chi_{2\text{eff}}} = \langle \dfrac{\chi_1}{1+\chi_2} \rangle$ <br> $\dfrac{1}{1+\chi_{2\text{eff}}} = \langle \dfrac{1}{1+\chi_2} \rangle$ |
| $\omega_{01}, a_2$ | $\omega_{01}\downarrow, \omega_{02}\uparrow$ | $1/\chi_{1\text{eff}} = \langle 1/\chi_1 \rangle$ <br> $\dfrac{1}{1+\chi_{2\text{eff}}} = \langle \dfrac{1}{1+\chi_2} \rangle$ |
| $\omega_{01}, \omega_{02}$ | $\omega_{01}\downarrow, \omega_{02}\uparrow$ | $1/\chi_{1\text{eff}} = \langle 1/\chi_1 \rangle$ <br> $\dfrac{1}{1+\chi_{2\text{eff}}} = \langle \dfrac{1}{1+\chi_2} \rangle$ |
| $a_1, a_2$ | $\omega_{01}\uparrow, \omega_{02}\uparrow$ | $\dfrac{1}{1+\chi_{1\text{eff}}+\chi_{2\text{eff}}} = \langle \dfrac{1}{1+\chi_1+\chi_2} \rangle$ |

Table 1 Summary of effective medium formula for time-modulated acoustic metamaterials with two resonances for different modulation cases: $\beta_{\text{eff}} = 1 + \chi_{1\text{eff}} + \chi_{2\text{eff}}$.

**Acknowledgements**

J.L. acknowledges support from Hong Kong Research Grants Council (RGC) grant (16303019) and AoE/P-502/20.

**Author contributions** J. Li conceived the idea of time-varying metamaterials with multiple resonances. X. Z. and J. Li set up the theoretical model and the numerical analysis. J. Li and H. W. provided insightful comments on the theoretical explanation. All authors contributed to scientific discussions of the results, explanations and wrote the manuscript.

**Competing interests**

Authors declare no competing interests.

**Data availability**

The datasets used and analyzed during the current study are available from the corresponding author upon reasonable request.